# The Einstein nanocrystal

Dalía S. Bertoldi[(1)]
Armando Fernández Guillermet[(1)(2)]
Enrique N. Miranda[(1)(3)*]

(1)Facultad de Ciencias Exactas y Naturales
Universidad Nacional de Cuyo, Mendoza, Argentina
(2)Instituto Balseiro
Comisión Nacional de Energía Atómica, Bariloche, Argentina
(3)IANIGLA
CONICET, CCT-Mendoza, Mendoza, Argentina

(*)E-mail: emiranda@mendoza-conicet.gov.ar

**Abstract**
We study the simplest possible model of nanocrystal consisting in a simple cubic lattice with a small number of atoms ($N_A \sim 10\text{-}10^3$), where each atom is linked to its nearest neighbor by a quantum harmonic potential. Some properties (entropy, temperature, specific heat) of the nanocrystal are calculated numerically but exactly within the framework of the microcanonical ensemble. We find that the presence of a surface in the nanocrystal modifies the thermostatistic properties to a greater extent than the small number of atoms in the system. The specific heat $C_v$ behaves similarly to the Einstein solid, with an asymptotic value for high temperatures that differs from that of the Dulong-Petit law by a term of the order of $N_A^{-1/3}$ and that can be explained easily in terms of the surface. The entropy is non-additive, but this is due to the presence of the surface and we show that the additivity is recovered in the thermodynamic limit. Finally, we find that, when calculations follow the canonical ensemble, results differ little for small systems ($N_A = 27$) and are inexistent for larger systems ($N_A = 1000$).

# Introduction

Due to the current boom of physics at a nanoscale, in this work we study a simple nanocrystal model inspired in the well-known Einstein model of solids [1, 2, 3, 4]. The use of highly idealized models is a common practice in physics because, even though they do not allow for precise quantitative predictions, their simplicity enables a clear qualitative understanding of a problem. The nanocrystal model proposed can be analyzed exactly, thus showing clearly which factors contribute to its differing behavior from that of the macroscopic solid. In a previous paper [5] we analyzed the Einstein solid and the two-level system with few elements (10-100); here, we want to analyze the effect of a surface. Therefore, we first need to compare the few-particle effect and the surface effect. Also, certain conceptual concerns should be taken into account, such as the additivity of the entropy and the equivalence between results obtained with the microcanonical and canonical ensembles. It is known that both ensembles lead to the same results in the thermodynamic limit, but not when dealing with finite systems. An introduction to this issue can be found in [6] and suitable teaching examples in [5].

This article continues in Section 2, which introduces the model and the equations that will be used throughout this paper. Section 3 presents the results: first we compare the few-particle and surface effects; then, we show how the specific heat converges to the value of the thermodynamic limit when the nanocrystal size is increased; the additivity of the entropy is dealt with next; and finally, we show how the results of the microcanonical ensemble differ from those of the canonical ensemble. Last, Section 4 summarizes the results.

## 1. Model

The Einstein solid, a well-known example for any advanced undergraduate student [1, 2, 3, 4], is the starting point of our model. We intend to model a nanocrystal, i.e. a crystal formed by a few tens or hundreds of atoms. Therefore, we assume it is a simple cubic lattice with $n_L$ atoms per side, spaced by a distance $a_0$. Each atom is linked to its nearest neighbors through a quantum "spring" (oscillator) –see Figure 1. It is assumed that all the oscillators have the same characteristic frequency $\upsilon_0$. Then, the total number of atoms $N_A$, the volume $V$ of the nanocrystal and the number of oscillators $N$ are:

$$N_A = n_L^3 \tag{1a}$$

$$V = (n_L - 1)a_0^3 \tag{1b}$$

$$N = 3(n_L^3 - n_L^2) \tag{1c}$$

It should be noted that the inner and surface atoms have a different number of neighbors, so that the number of quantum oscillators is not simply $3N_A$. A little thought shows that equation (1c) gives the right number of oscillators. When dealing with the Einstein solid, textbooks are interested in the thermodynamic limit and, therefore, the surface is not considered and the number of oscillators is simply $3N_A$.

The calculations will follow the microcanonical ensemble. The reason for this is that the nanocrystal is considered to be isolated and its energy constant. Using the canonical ensemble, the system would be connected to a reservoir at a constant temperature. However, it should be noted that, given the size of the nanocrystal, such interaction would get the system continuously out of equilibrium. Let us imagine the nanocrystal is in contact with a gas at room temperature: each impact of a gas molecule with the few-hundred-atom nanocrystal is a major event that would keep the system permanently out of equilibrium. This is why the proper way to study the nanocrystal is within the microcanonical ensemble.

The frequency of each oscillator is $v_0$, thus, the entropy $S$ of a cubic crystal with $N$ oscillators and a total energy $E$ energy is [1, 3,4]:

$$S(N,E) = k_B \ln\left[(N - 1 + E/hv_0)!/((N - 1)!\,(E/hv_0)!)\right] \tag{2}$$

Notice that $(E/hv_0)$ is the total number of energy quanta and it is an integer $M$. A detailed derivation of equation (2) is a standard topic in any statistical mechanics course. The usual trick is to evaluate how many different ways there are to put $M$ balls in $N$ boxes and the problem can be reduced to calculate the number of permutations of *(N-1+M)* objects (*M* balls and *N-1* rods) taking into account that balls (rods) are indistinguishable. The usual treatment of the problem in textbooks considers $N = 3N_A$, uses Stirling's approximation with factorials, and then applies the usual thermodynamic equations for temperature ($1/T = \partial S/\partial E$) and specific heat ($C_v = \partial E/\partial T$). However, a previous paper [5] showed that, when working with few particles, factorials must be treated exactly and derivatives replaced by finite differences to account for the discrete nature of energy. On one side, Stirling's approximation is valid for $N$ large but it is not obvious that it is also right for small systems. For this reason it is necessary to calculate the factorial exactly. On the other side, a central result of quantum physics is usually forgotten and the energy of a system of oscillators is taken as a continuous variable. Such assumption is incorrect since it is known since the early twentieth century that the energy of an oscillator is quantized, that is, there are $M$, $M + 1$,.. etc. energy quanta $M$ being an integer given by $M = E / h\upsilon_0$. Strictly speaking one should deal with finite differences, then $\partial S \to \Delta S = S(N, M+1) - S(N, M)$ y $\partial E \to \Delta E = (M+1) - M = 1$. Consequently, the temperature $T$ and the specific heat $C_v$ need to be computed as follows:

$$T = (\partial S/\partial E)^{-1} \to T(N,M) = [S(N,M+1) - S(N,M)]]^{-1} \tag{3a}$$

$$C_v = \partial E/\partial T \to C_v(N,M) = [T(N,M+1) - T(N,M)]^{-1} \tag{3b}$$

Note that (3a) and (3b) are not approximations, but represent reality; only when working with large numbers for $M$, the discrete nature of energy can be disregarded, taking energy as a continuous variable and using derivatives.

One improvement of the model would be to consider different frequencies for the oscillators on the surface and for those inside the crystal. This assumption certainly makes sense since it remarks the difference between the surface and the volume inside the crytal; however this modified model cannot be studied in the microcanonical ensemble. Suppose that the surface oscillator frequency is $\upsilon'$. Therefore two kinds of energy quanta should be considered: the $M'$ quanta with energy $h\upsilon'$ and the $M$ ones with energy $h\upsilon_0$. In the microcanonical formalism the system is isolated and $M$ and $M'$ are constant, consequently the volume and surface temperatures would be different but the system cannot reach thermal equilibrium since the energy quanta of the surface oscillators are not equivalent to those inside the crystal. If one is interested in analyzing this modified model, the canonical formalism has to be used: the nanocrystal is in contact with a heat reservoir that provides energy quanta of both frequencies and the surface and the inside volume can reach the same temperature.

## 2. Results

### 2.1. Few particle (FP) and finite size (FS) effects

We must distinguish now between two types of effects occurring when doing statistical mechanics with few elements. We analyzed a few-particle model in [5] without taking into account the surface, so that the number of oscillators was considered to be simply three times the number of atoms ($N = 3N_A$). The difference between these results and those obtained in the thermodynamic limit are due exclusively to the small number of particles in the system, so accordingly we will call this the few-particle (FP) effect. Now that we are considering the surface, the number of oscillators must be carefully recorded and, for this, the value for $N$ is that given by equation (1c). Thus, we can speak of the finite-size (FS) effect, which includes the few-particle the surface effects. The first question that arises is, then, which effect is more relevant: that due to the few particles or the one due to the presence of a surface?

To answer this question, we evaluated numerically but exactly, i.e. without any approximations, the entropy given by equation (2). In one case, we consider $N=3N_A$ and obtain an entropy $S^{FP}$ that only includes the few-particle effect. In the other case, the number of oscillators $N$ is taken from equation (1c) and we obtain an entropy $S^{FS}$ that includes both the few particles and the surface. The difference between both entropies gives the surface effect. The reference value is the entropy $S^{th}$ of the classical Einstein solid in the thermodynamic limit [1, 4]:

$$S^{th}(3N_A, E) = 3N_A k_B \ln\left[1 + \frac{E}{3N_A h\nu_0}\right] + \frac{E}{h\nu_0} k_B \ln\left[1 + \frac{3N_A h\nu_0}{E}\right] \tag{4}$$

Once the entropies $S^{FP}$ and $S^{FS}$ are known, it is possible to find the temperatures $T^{FP}$ and $T^{FS}$, and the specific heats $C_v^{FP}$ and $C_v^{FS}$ using equations (3a) and (3b). It is interesting to compare these results to those of the classical Einstein solid in the thermodynamic limit [1, 2, 3, 4]. According to the notation used in this article, $T^{th}$ and $C^{th}$ turn out to be:

$$T^{th}(3N_A, E) = \frac{h\nu_0}{k_B}\left[\ln\left(1 + \frac{3N_A h\nu_0}{E}\right)\right]^{-1} \tag{6}$$

$$C_v^{th}(3N_A, E) = \frac{k_B}{3N_A}\left(\frac{E}{h\nu_0}\right)^2 \left(1 + \frac{3N_A h\nu_0}{E}\right)\left[\ln\left(1 + \frac{3N_A h\nu_0}{E}\right)\right]^2 \tag{7}$$

Now, we can define the relative effect for each of the thermodynamic quantities. If $X$ represents the quantity of interest ($S$, $T$, $C_v$), the relative effect associated with the few particles ($\Delta X^{FP}$) and that associated with the finite size ($\Delta X^{FS}$) can be defined as follows:

$$\Delta X^{FP} = 100 * (X^{FP} - X^{th})/X^{th} \tag{5a}$$

$$\Delta X^{FS} = 100 * (X^{FS} - X^{th})/X^{th} \tag{5b}$$

Thus, it is clear which is the impact due to the small number of particles and that due to the surface. The results can be seen in Figure 2.

Figure 2a shows the difference in entropy for two sizes of the few-particle system ($n_L = 3$ and $n_L = 10$). Consistently with the findings of [5], there is no discernible difference between the exact value and that of the thermodynamic limit for the system with *1000* atoms. On the contrary, Figure 2b considers the surface and its effect is considerable in the small system ($\Delta S^{FS} \sim 22\%$) as well as in the large system ($\Delta S^{FS} \sim 5\%$). This demonstrates the impact of the presence of the surface in the thermodynamics of the nanocrystal.

The question arises of how this effect is shown in experimentally observable quantities. The answer is in Figures 2c and 2d. Figure 2c shows the impact on the specific heat due to the few particles effect. For the system with *1000* atoms, the specific heat is not different from the thermodynamic value. On the contrary, when considering the surface – Figure 2d–, the specific heat is lower ($\Delta C_v^{FS}$ ~*10%*) than the thermodynamic value, even for the larger system. For the system with $n_L = 3$, the difference between the specific heat and the thermodynamic value is even greater ($\Delta C_v^{FS}$ ~*30%*).

In conclusion, the surface plays a much more important role that the finite number of particles in the thermophysical properties of the nanocrystal.

One may ask whether it makes sense to evaluate the properties of a system using the exact approach used in this work. The answer is: it depends on the size of the system under study. For atomic clusters with tens of atoms, the calculation should be made exactly with the microcanonical formalism as there is a noticeable difference with the thermodynamic values. Conversely, if one has a system with thousands of atoms, the conventional results obtained with the canonical formalism can be used.

### 2.2. Specific heat scaling

Figure 3a illustrates the atomic specific heat of the nanocrystal as a function of temperature for the different sizes, and it also shows the curve of the usual Einstein solid. It is clear that the nanocrystal behavior is qualitatively similar to that of the macroscopic solid: with high temperatures, the specific heat reaches a constant value approaching the Dulong-Petit law. It should be highlighted that, in the units used in this article, $C_v/N_A$ goes to *3* for the solid in the thermodynamic limit. Quantitatively, the specific heat of the nanocrystal is lower than that of the solid. It is interesting to illustrate the asymptotic value reached by the $C_v$ of the nanocrystal as a function of its size. As seen in Figure 3b, it can be inferred the following scaling relationship:

$$C_v/N_A = 3 - 3N_A^{-1/3} \tag{6}$$

This scaling is caused by the presence of the surface in the crystal. In fact, equations (1a) and (1c) give that the number of oscillators can be written as $N/N_A = 3 - 3N_A^{-1/3}$. The surface atoms are associated with fewer oscillators than the inner atoms, and those missing oscillators account for the difference in the specific heat of the solid and the nanocrystal.

### 2.3. Non-additivity of the entropy

The entropy in the thermodynamic limit is an extensive property; however, this may not be so at a microscopic scale since the range of interactions is comparable to the size of the system. To verify if the entropy is additive or not in our nanocrystal model, we carried out the following numeric experiment –see Figure 4a–: we took a nanocrystal $B$ of side $2n_L$ (atoms per side) and divided into 8 nanocrystals $A$ of side $n_L$. Similarly, the energy quanta $M$ of the original nanocrystal were distributed evenly among the $8$ new nanocrystals. The question is whether the initial and final entropies are equal; and the answer is no. We find that $S(B) > 8\ S(A)$. Figure 4b shows the difference in entropy $\delta S = S(B) - 8\ S(A)$ as a function of the size of the system. It is clear that the difference increases as $N_A^{2/3}$. Naturally, what makes sense is the entropy per atom and, therefore, $\delta S / N_A \sim N_A^{-1/3}$ reaches zero in the thermodynamic limit.

Fitting numerically the straight line shown in Figure 5, results in a slope of approximately $12$ which can be understood easily. The difference in the number of oscillators $\delta N_{osc}$ between the large cube and the 8 small ones is:

$$\delta N_{osc} = [3((2n_L)^3 - (2n_L)^2)] - 8[3((n_L)^3 - (n_L)^2)]$$

$$= 12 n_L^2$$

$$= 12 N_A^{2/3} \qquad (7)$$

The non-additivity of entropy is generated by the partition of the initial cube into 8 smaller cubes, which creates a larger surface and, therefore, increases the number of surface atoms with fewer oscillators associated than the inner atoms.

### 2.4. Microcanonical and canonical ensembles

The calculations performed so far have followed the microcanonical ensemble. As stated above, this is the proper method for analyzing systems with a few hundred atoms. If a system of this size is connected to an energy reservoir, their interaction would get the nanocrystal out of equilibrium. Besides, it is usually claimed in any statistical mechanics course that the results obtained using either the canonical or microcanonical ensembles coincide in the thermodynamic limit. The question that arises is, then, whether this also happens when working with very few particles.

In our notation, $Z^{FS}$ is the partition function of the nanocrystal of side $n_L$ and with $3(n_L^3 - n_L^2)$ oscillators, whereas $Z^{FP}$ is the partition function of a set of $3n_L^3$ oscillators. The Einstein temperature is defined as: $\Theta = h\nu_0 / k_B$ and considering that the partition function $z$ of a quantum oscillator is: $z = (1 - e^{-\Theta/T})^{-1}$, it is a simple exercise of statistical mechanics to obtain $Z^{FP}$ and $Z^{FS}$, and from these to compute the specific heats:

$$\frac{C_v^{can\ FP}}{k_B N_A} = 3\left(\frac{\Theta}{T}\right)^2 \frac{e^{-\Theta/T}}{(1-e^{-\Theta/T})^2} \qquad (8)$$

$$\frac{C_v^{can\ FS}}{k_B N_A} = 3\left(1 - \frac{1}{n_L}\right)\left(\frac{\Theta}{T}\right)^2 \frac{e^{-\Theta/T}}{(1-e^{-\Theta/T})^2} \tag{9}$$

In equations (8) and (9), the superscript "*can*" indicates that the specific heats are obtained using the canonical ensemble –in contrast to our work so far following the microcanonical ensemble. By comparing (8) and (9), it can be seen that the surface introduces a correction of *$1/n_L$* to the specific heat.

Do the specific heat values computed with both ensembles coincide? The answer is in Figure 5 and it shows a very good agreement between both results, even for the small system ($n_L = 3$). The specific heats graphed have been calculated within the microcanonical (superscript *mic*) and canonical (superscript *can*) ensembles, for the FP and FS cases. The energy per atom ($M/N_A$), which is the natural variable in the microcanonical ensemble, was used as independent variable. For the case of the specific heats (8) and (9), equation (6) allows to replace the temperature with the energy per atom. For the case of $n_L = 10$, there is no discernible difference between the results from the canonical and microcanonical ensembles and has, therefore, not been illustrated.

## 3. Conclusions

As in [5], the first point to highlight is the strength of the results in statistical mechanics: even with systems formed by 27 elements, the thermostatistical properties are not essentially different from the results found in the thermodynamic limit. In this sense, the usual approximations of statistical physics are reassuring; one may trust them.

Related to this point, the surface is clearly important for the properties of the studied systems. For the system with 27 atoms, when there is no consideration for the surface, the specific heat differs by approximately *1%* from the thermodynamic value, while there is no difference for the system with 1000 atoms. However, when the surface is considered, the difference is ~*30%* in the first case and ~*10%* in the second. This means that the presence of the surface has a greater impact than doing statistical mechanics with few particles.

It is clear for what sizes one has to use the exact microcanonical calculation and when the usual thermodynamic results can be employed. For systems formed by dozens of particles, it is justified to proceed in the meticulous way followed in this article. For systems with several thousand atoms, usual thermodynamic results are good enough.

The entropy of the nanocrystal is non-additive, apparently contrary to a basic principle of thermodynamics. However, this is not so; the non-additivity can be explained by the presence of the surface, and it can be clearly shown that it is recovered in the thermodynamic limit.

Finally, our work confirms that the results obtained in the canonical and microcanonical ensembles are equal for systems with $N_A > 10^3$. This is relevant because calculations are always easier in the canonical ensemble. Thus, people working with nano-systems can use the calculation techniques of the canonical ensemble even when the problem under analysis is isolated and calls for the microcanonical ensemble.

In a nutshell: one can use statistical mechanics even for systems with tens of particles but should be aware of the effects of the surface.

Financial support from Universidad Nacional de Cuyo through Proyecto SECTyP 06/M072 is acknowledged. One of the authors (ENM) thanks Diego Molina for helping in the last stage of this project.

# Figure captions

**Figure 1:**

The Einstein nanocrystal analyzed in this article is a simple cubic structure with $n_L$ atoms per side (in this case, $n_L = 3$). The spacing between atoms is $a_0$. Atoms interact with each other through a quantum harmonic potential with a frequency $v_0$, which is the same for all oscillators. The total number of atoms is $N_A = n_L^3$ while the number of oscillators is $N = 3(n_L^3 - n_L^2)$. The fact that surface atoms are associated to fewer oscillators than the inner atoms, have measurable consequences on the thermophysical properties of the crystal.

**Figure 2:**

This figure shows how the nanocrystal is affected by having few particles (FP) –figures (a) and (b)– and a finite size (FS) including the small number of particles and the surface – figures (c) and (d). The entropy –(a) and (c)– and the specific heat –(b) and (d)– have been calculated exactly, according to the equations of the microcanonical ensemble, and the relative difference between those results and the thermodynamic limit values have been evaluated for the FP and FS cases. Results are expressed in relation to the energy per atom ($M/N_A$). In the FP case, the exact results coincide with the thermodynamic values for the large system ($n_L = 10$), except at very low energies. Conversely, in the FS case, the difference with the thermodynamic values is significant even for the large system. This demonstrates that the presence of the surface affects the system properties more than the small number of particles.

**Figure 3:**

(a) The specific heat of the nanocrystal, evaluated exactly within the microcanonical ensemble, is plotted as a function of the temperature for the different sizes. The curve corresponding to the thermodynamic limit has also been included for reference. It can be observed that the nanocrystal behaves qualitatively similarly to the solid and that the specific heat reaches an asymptotic value approaching the Dulong-Petit law.

(b) The asymptotic value of the specific heat is shown in terms of $N_A^{-1/3}$. It is clear that the specific heat converges to the value of the Dulong-Petit law in the thermodynamic limit. Its special relation with $N_A$ is explained in the text.

**Figure 4:**

(a) This figure illustrates the numerical experiment performed to verify the additivity of the entropy: a cube of side $2n_L$ is partitioned into $8$ small cubes of side $n_L$. The question is whether the addition of the entropies of the small cubes is equal to the entropy of the large cube.

(b) This shows the difference in entropy between the large cube and the 8 small ones. It can be seen that the difference increases as $N_A^{2/3}$. Clearly, the entropy is non-additive, which can be explained by the effect of the surface. Besides, the quantity to take into account is the entropy per atom and, in that case, the difference of entropies is expressed $N_A^{-1/3}$, and it is consequently zero in the thermodynamic limit.

**Figure 5:**

The specific heat is shown in relation to the energy per atom for a nanocrystal with $n_L = 3$. It has been calculated exactly in the microcanonical (superscript *mic*) and canonical (superscript *can*) ensembles, for the FP and FS cases. As it can be seen, both ensembles lead to similar results, except at low energies. The presence of the surface has a much greater impact than the ensemble chosen for the calculations.

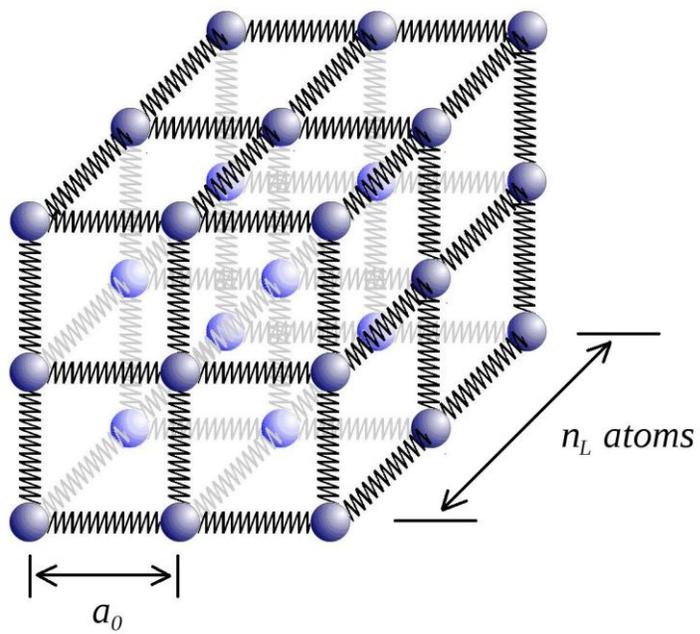

**Figure 1**

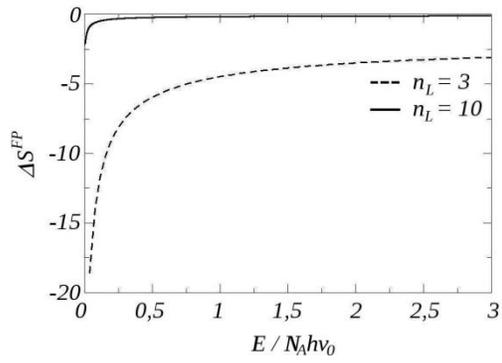 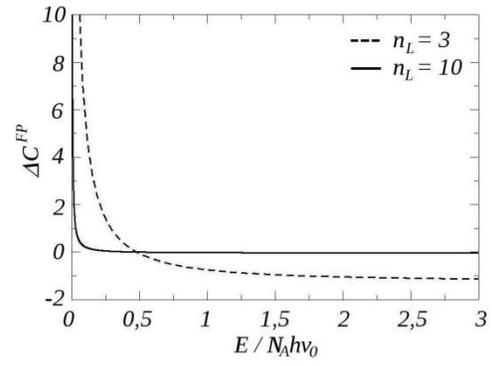

**(a)**  **(b)**

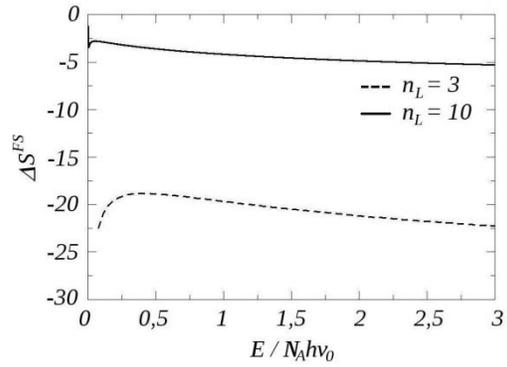 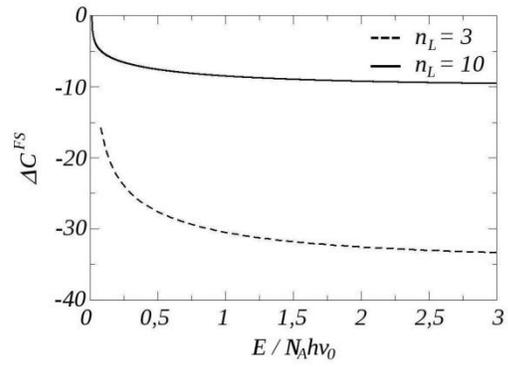

**(c)**  **(d)**

**Figure 2**

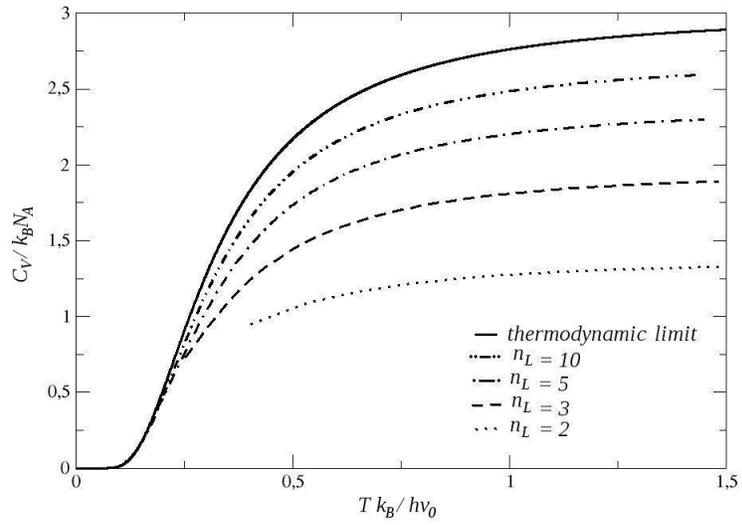

**(a)**

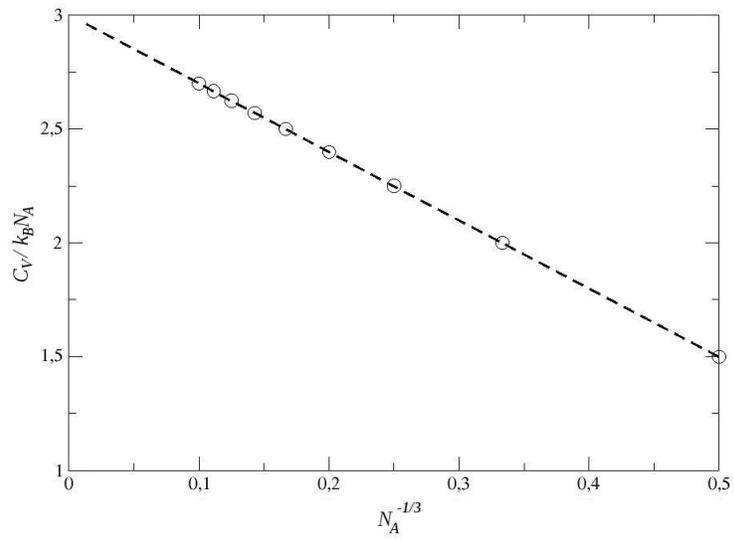

**(b)**

**Figure 3**

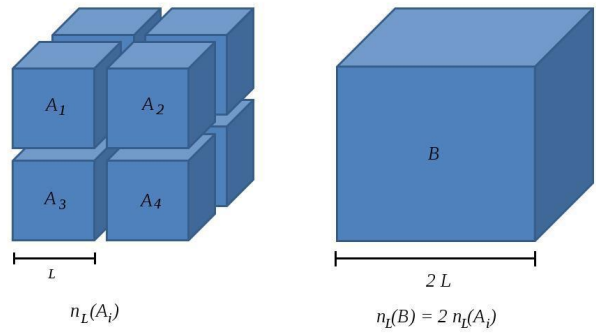

$n_L(A_i)$

$n_L(B) = 2\, n_L(A_i)$

**(a)**

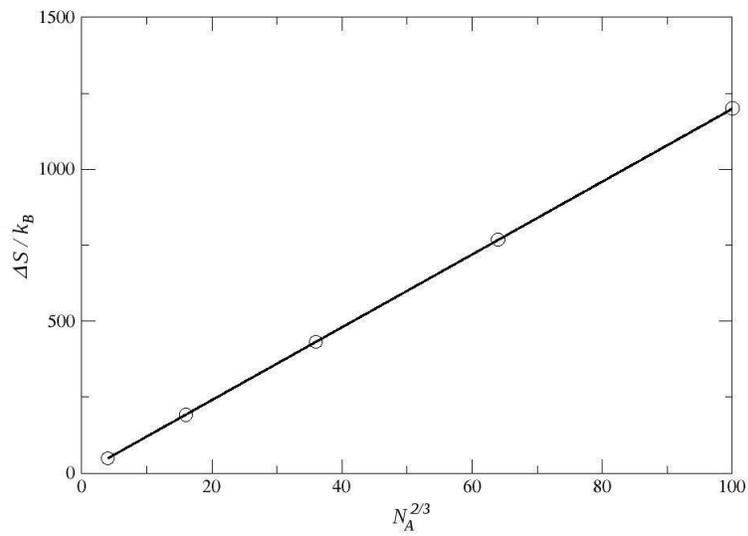

**(b)**

**Figure 4**

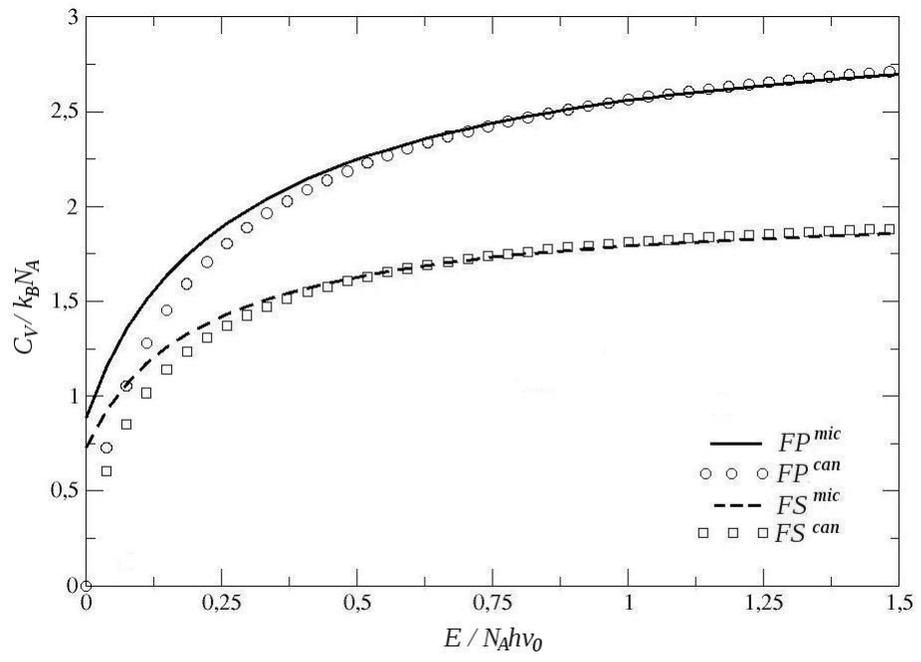

**Figure 5**